\documentclass[twoside,leqno,twocolumn]{article}
\usepackage[letterpaper]{geometry}

\usepackage{siamproceedings}

\usepackage[T1]{fontenc}
\usepackage{amsfonts}
\usepackage{graphicx}
\usepackage{epstopdf}
\usepackage{enumitem}
\usepackage{hyperref}  
\usepackage{algorithm}
\usepackage{algpseudocodex}
\usepackage{xcolor}
\usepackage{multirow}
\usepackage{tikz}
\usepackage{booktabs}
\usepackage{amssymb}
\ifpdf
  \DeclareGraphicsExtensions{.eps,.pdf,.png,.jpg}
\else
  \DeclareGraphicsExtensions{.eps}
\fi

\usepackage{amsopn}

\algnewcommand\algorithmicparfor{\textbf{parallel for}}%
\algnewcommand\algorithmicendparfor{\textbf{end parallel for}}%
\makeatletter
\algdef{SE}[PARFOR]{ParFor}{EndParFor}[1]{\algpx@startIndent\algpx@startCodeCommand\algorithmicparfor\ #1\ \algorithmicdo}{\algpx@endIndent\algpx@startCodeCommand\ algorithmicendparfor\ \ algorithmicparfor}%
\ifbool{algpx@noEnd}{%
  \algtext*{EndParFor}%
  \apptocmd{\EndParFor}{\algpx@endIndent}{}{}%
}{}%
\pretocmd{\ParFor}{\algpx@endCodeCommand}{}{}
\ifbool{algpx@noEnd}{%
  \pretocmd{\EndParFor}{\algpx@endCodeCommand[1]}{}{}%
}{%
  \pretocmd{\EndParFor}{\algpx@endCodeCommand[0]}{}{}%
}%
\makeatother

\begin{document}

\title{\Large Parallelizing the Approximate Minimum Degree Ordering Algorithm:\\ Strategies and Evaluation\thanks{The U.S. Government retains a nonexclusive, royalty-free license to publish or reproduce the published form of this contribution, or allow others to do so, for U.S. Government purposes.}}
\author{Yen-Hsiang Chang\thanks{University of California, Berkeley}
\and Ayd{\i}n Bulu\c{c}\thanks{NVIDIA Corporation} \thanks{work done while employed at Lawrence Berkeley National Laboratory and University of California, Berkeley}   
    \and James Demmel\footnotemark[2] 
}

\date{}

\maketitle

\fancyfoot[R]{\scriptsize{Copyright \textcopyright\ 2026 \\
Copyright for this paper is retained by authors}}

\begin{abstract}
  The approximate minimum degree algorithm is widely used before numerical factorization to reduce fill-in for sparse matrices. While considerable attention has been given to the numerical factorization process, less focus has been placed on parallelizing the approximate minimum degree algorithm itself. In this paper, we explore different parallelization strategies, and introduce a novel parallel framework that leverages multiple elimination on distance-2 independent sets. Our evaluation shows that parallelism within individual elimination steps is limited due to low computational workload and significant memory contention. In contrast, our proposed framework overcomes these challenges by parallelizing the work across elimination steps. To the best of our knowledge, our implementation is the first scalable shared memory implementation of the approximate minimum degree algorithm. Experimental results show that we achieve up to a 7.29$\times$ speedup using 64 threads over the state-of-the-art sequential implementation in SuiteSparse.
\end{abstract}
\section{Introduction.}
Efficiently solving sparse linear systems is a cornerstone of many scientific and engineering applications.
Cholesky factorization, a widely used direct method for solving symmetric positive definite systems, often encounters challenges when the given matrix $A$ is sparse.
In such cases, zeros in $A$ can transform into nonzeros, known as fill-ins, in the Cholesky factor.
These fill-ins can significantly increase both memory usage and computational cost, undermining the efficiency of the factorization.
To address this, fill-reducing orderings seek to find a permutation matrix $P$ such that the permuted matrix $PAP^T = LL^T$ reduces the number of fill-ins in the Cholesky factor $L$.
However, minimizing the number of fill-ins is an NP-complete problem~\cite{yannakakis1981computing}, and as a result, various heuristics have been developed, such as nested dissection~\cite{george1973nested, gilbert1986analysis} and approximate minimum degree (AMD)~\cite{amestoy1996approximate}.
This paper solely focuses on parallelizing the AMD algorithm in shared memory since to the best of our knowledge, no parallel implementation currently exists.

The need for a scalable shared memory parallel AMD implementation is motivated by literature and practical use cases.
Trotter et al.~\cite{trotter2023bringing} recently conducted an extensive study on 490 matrices and found that while nested dissection holds some advantage over AMD in reducing fill-ins, the ordering time for nested dissection is up to an order of magnitude longer than that of AMD. 
Furthermore, while significant work has been done to parallelize nested dissection~\cite{chevalier2008pt,karypis1998parallel,lasalle2015efficient}, no similar efforts have been made for the AMD algorithm, highlighting a key gap in the literature.
Meanwhile, as shown in Table~\ref{tab:motivation}, GPU Cholesky solvers have steadily improved over time.
In contrast, the standalone sequential AMD implementation on CPU is poised to become a bottleneck as other components of the solver continue to improve.
To make the AMD algorithm more future-proof, our goal is therefore to design a scalable shared memory AMD implementation, closing the gap in theory and in practice.

\begin{table}[htbp]
  \vspace{-1em}
  \caption{Time (in seconds) for the sequential AMD from SuiteSparse (v7.12.1)~\cite{suitesparsegithub} on an AMD EPYC 7763 CPU, and for Cholesky solver cuSolverSp~\cite{cusolver} (v11.6.1) and its successor cuDSS~\cite{cudss} (v0.7.1) to solve the reordered system on an A100 GPU in double precision.}
  \label{tab:motivation}
  \centering
  \begin{tabular}{l|c|cc}
    \toprule
    Matrix & AMD & cuSolverSp & cuDSS \\
    \midrule
    \texttt{nd24k} & 0.82 & 117.17 & 1.97 \\
    \texttt{ldoor} & 1.42 & 11.01 & 3.03 \\
    \texttt{Flan\_1565} & 4.85 & out of memory & 18.92 \\
    \texttt{Cube5317k} & 13.94 & out of memory & 43.90 \\
    \bottomrule 
  \end{tabular}
\end{table}

Parallelizing the AMD algorithm presents two main challenges.
First, each elimination step involves selecting a pivot vertex with minimum approximate degree, eliminating it, and then updating its neighborhood and the approximate degrees of the neighbors.
These steps are inherently sequential, as a new pivot cannot be selected until all updates from the previous pivot are complete.
While neighborhood and approximate degree updates could, in theory, be parallelized, the opportunities for efficient parallelism are limited due to little amount of work and high memory contention.
Second, although the multiple elimination strategy from the sequential multiple minimum degree algorithm~\cite{liu1985modification} can help reduce dependencies by selecting a maximal independent set of pivots, this approach introduces significant overlap in these pivots' neighborhoods.
While such overlap is beneficial in the sequential setting, it introduces contention on the underlying graph structure and complicates the maintenance of approximate degrees, ultimately hindering scalability when attempting to parallelize the AMD algorithm using multiple elimination.

In this paper, we address these challenges by selecting distance-2 independent sets of pivots, enabling parallel elimination steps while keeping approximate degree updates simple and efficient.
We further exploit key properties of the AMD algorithm to design specialized concurrent data structures---including a concurrent graph representation and concurrent approximate degree lists---that eliminate the need for dynamic memory allocation and significantly reduce memory contention.
Together, these optimizations lead to the first scalable implementation of the AMD algorithm, outperforming the long-standing sequential version in SuiteSparse~\cite{suitesparsegithub} with up to a 7.29$\times$ speedup using 64 threads.
\section{Preliminaries.}
\begin{figure*}[t]
    \centering
    \includegraphics[width=\linewidth]{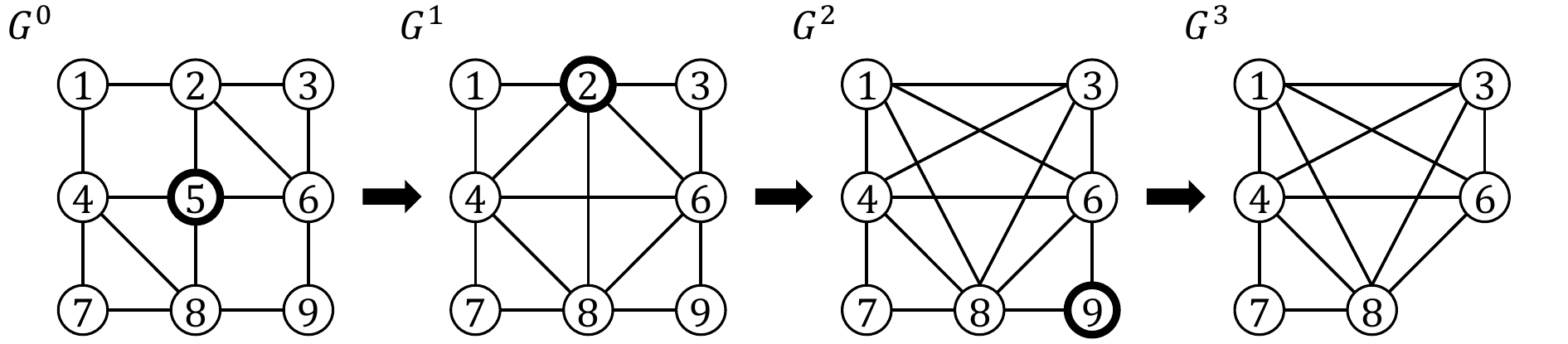}
    \vspace{-1em}
    \caption{An example illustrating how elimination graphs work. For demonstration purposes, we eliminate vertices 5, 2, and 9 in order, rather than following the minimum degree criterion. When a vertex is eliminated, its neighbors form a clique.}
    \label{fig:elimination}
\end{figure*}

\begin{figure*}[t]
    \centering
    \includegraphics[width=\linewidth]{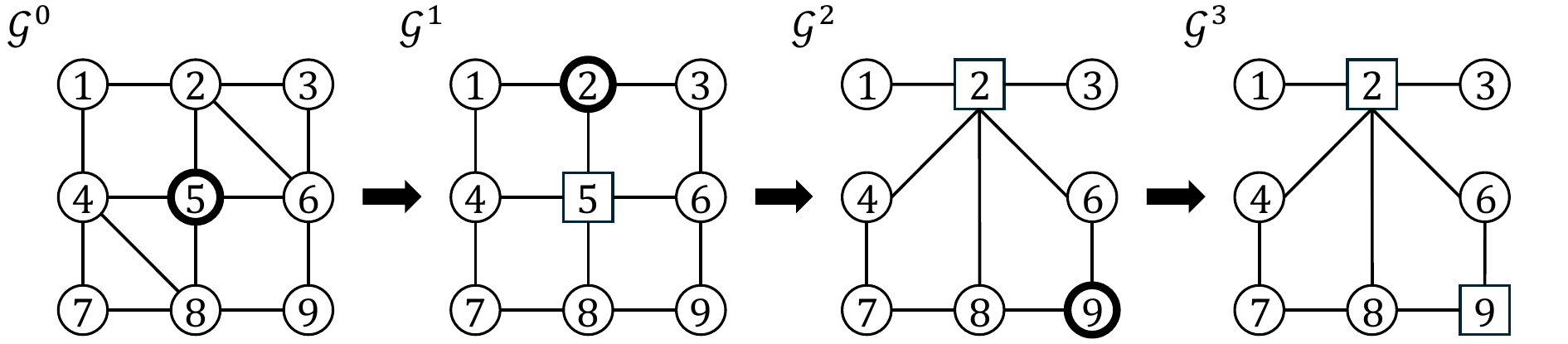}
    \vspace{-1em}
    \caption{An example illustrating how quotient graphs work. For demonstration purposes, we eliminate variables 5, 2, and 9 in order, rather than following the minimum degree criterion. Circles represent variables and squares represent elements. When variable 2 is eliminated in $\mathcal{G}^1$, its neighborhood in $G^1$ is recovered using the clique encoded on element 5 and the remaining original neighbors, forming a new clique on variables 1, 3, 4, 6, and 8.}
    \label{fig:quotient}
\end{figure*}

Before reviewing the approximate minimum degree (AMD) algorithm~\cite{amestoy1996approximate}, we begin with its predecessors, the minimum degree algorithm~\cite{rose1972graph} and the multiple minimum degree algorithm~\cite{liu1985modification}.
From this point onward, we disregard the positive definiteness of the matrices, as it does not impact the ordering algorithms discussed in this paper.

\begin{table}[h]
\centering
\caption{Summary of notation used.}
\label{tab:notation}
\begin{tabular}{cl}
\hline
Symbol & Description \\
\hline
$A$ & An $n \times n $ sparse symmetric matrix \\
$p$ & The selected pivot \\ 
$G^k$ & The $k$th elimination graph \\
$\mathcal{G}^k$ & The $k$th quotient graph \\
\multirow{2}{*}{$V^k$} & The set of vertices in $G^{k}$, equivalent to \\
& the set of variables in $\mathcal{G}^k$ \\
$E^k$ & The set of edges in $G^k$ \\
$\overline{V}^k$ & The set of elements in $\mathcal{G}^k$ \\
$N^k_v$ & The neighborhood of vertex $v$ in $G^k$ \\
$\mathcal{A}^k$ & The connections among $V^k$ in $\mathcal{G}^k$ \\ 
$\mathcal{E}^k$ & The connections from $V^k$ to $\overline{V}^k$ in $\mathcal{G}^k$ \\ 
$\mathcal{L}^k$ & The connections from $\overline{V}^k$ to  $V^k$ in $\mathcal{G}^k$ \\
$\mathcal{A}^k_v$ & The variables adjacent to variable $v$ in $\mathcal{G}^k$ \\
$\mathcal{E}^k_v$ & The elements adjacent to variable $v$ in $\mathcal{G}^k$ \\
$\mathcal{L}^k_e$ & The variables adjacent to element $e$ in $\mathcal{G}^k$ \\
\hline
\end{tabular}
\end{table}

\subsection{Minimum Degree Algorithm.}
The minimum degree algorithm is most naturally described using elimination graphs~\cite{rose1972graph}.
Let $A$ be an $n \times n$ symmetric sparse matrix.
The sparsity pattern of $A$ can be represented by a graph $G^0 = (V^0, E^0)$, where $V^0 = \{1, 2, \cdots, n\}$, and each vertex $i$ corresponds to row/column $i$ of $A$. 
An edge $(i, j)$ is in $E^0$ if and only if $A_{i, j} \neq 0$.
The \textit{elimination graph} $G^k = (V^k, E^k)$ represents the sparsity pattern of the remaining submatrix after the first $k$ pivots have been selected and eliminated during Cholesky factorization.
Let $N^k_v$ denote the neighborhood of vertex $v$ in the elimination graph $G^k$.
The elimination process begins with $G^0$.
To obtain $G^k$ from $G^{k - 1}$, the $k$th pivot $p$ is selected from $V^{k - 1}$, removed from the vertex set, and its neighbors are connected to form a clique---reflecting the fill-ins introduced by an outer product update in Cholesky factorization.
An example is presented in Figure~\ref{fig:elimination}.
Formally, the update can be expressed as:
\begin{align*}
    V^k &= V^{k - 1} \setminus \{p\}, \\
    E^k &= \left(V^k \times V^k\right) \cap \left(E^{k - 1} \cup \left(N^{k - 1}_p \times N^{k - 1}_p\right)\right).
\end{align*}
The minimum degree algorithm selects, at each step, the vertex with the smallest degree in $G^{k - 1}$ as the $k$th pivot.
Since multiple vertices can have the same degree, a tie-breaking strategy is required.
Ties are typically resolved arbitrarily or based on the initial ordering; however, it has been shown that the choice of tie-breaking strategy can significantly affect the number of fill-ins produced~\cite{george1989evolution}.

\subsection{Quotient Graphs.}
A key challenge with elimination graphs is that the storage needed for $G^k$ is not known in advance due to the formation of cliques during elimination, which necessitates dynamic memory management.
The quotient graph~\cite{george1980fast}, also known as the generalized element model or superelement~\cite{eisenstat1976applications, speelpenning1978generalized,  duff1983multifrontal}, addresses this issue by representing cliques more compactly---storing only the set of vertices in each clique.
George and Liu~\cite{george1981computer} showed that the quotient graph takes no more storage than the original graph.

Following historical terminology, we refer to eliminated vertices from the original graph as \textit{elements}, and those yet to be eliminated as \textit{variables}.
To stay consistent with the literature, we adopt the notations used in~\cite{amestoy1996approximate}.
The \textit{quotient graph} $\mathcal{G}^k = \left(V^k, \overline{V}^k, \mathcal{A}^k, \mathcal{E}^k, \mathcal{L}^k\right)$ provides an implicit representation of the elimination graph $G^k$. Here, $V^k$ denotes the set of variables, $\overline{V}^k$ the set of elements, $\mathcal{A}^k$ the set of variable-to-variable connections, $\mathcal{E}^k$ the connections from variables to elements, and $\mathcal{L}^k$ the connections from elements to variables.
Notably, there is no connection between elements, and while $\mathcal{E}^k$ and $\mathcal{L}^k$ are essentially transposes of each other, they are maintained separately in practical implementations for efficiency.
In essence, the variables $V^k$ correspond to the uneliminated vertices, with $\mathcal{A}^k$ being the remaining edges from the original graph; the elements $\overline{V}^k$ represent cliques formed during elimination, with their connections compactly encoded in $\mathcal{E}^k$ and $\mathcal{L}^k$. 

The elimination process begins with the initial quotient graph $\mathcal{G}^0 = (V^0, \emptyset, E^0, \emptyset, \emptyset)$.
For simplicity, we omit the superscript $k$ when the context is clear.
Given a variable $v$ and an element $e$ in the quotient graph, we define the following adjacency sets:
\begin{align*}
    \mathcal{A}_v &= \{v' \mid (v, v') \in \mathcal{A}\} \subseteq V, \\
    \mathcal{E}_v &= \{e' \mid (v, e') \in \mathcal{E}\} \subseteq \overline{V}, \\
    \mathcal{L}_e &= \{v' \mid (e, v') \in \mathcal{L}\} \subseteq V.
\end{align*}
The quotient graph and the elimination graph are closely related through the following invariant: for any vertex $v$ in $G$, which also appears as a variable in $\mathcal{G}$, its neighborhood is:
\begin{align}\label{eq:invariant}
    N_v = \left(\mathcal{A}_v \cup\bigcup_{e \in \mathcal{E}_v} \mathcal{L}_e\right)\setminus \{v\}.
\end{align}
In other words, the neighborhood of $v$ in the elimination graph can be reconstructed from the quotient graph by interpreting the remaining original neighbors and the clique information encoded in the adjacent elements.

When a variable $p$ is selected as the $k$th pivot from $\mathcal{G}^{k - 1}$, it becomes an element in $\mathcal{G}^k$.
Its neighborhood $N_p$ is computed using Equation~(\ref{eq:invariant}).
To preserve the invariant for the remaining variables, the new clique induced by $p$ is added to $\mathcal{L}$ encoded as $\{p\} \times N_p$, and similarly $N_p \times \{p\}$ is added to $\mathcal{E}$. 
Additionally, connections now covered by this newly formed clique are removed.
An example is presented in Figure~\ref{fig:quotient}.
Formally, the connection updates are defined as:
\begin{align*}
    \mathcal{A}^k &= \left(\mathcal{A}^{k - 1} \setminus \left(N_p^{k - 1} \times N_p^{k - 1}\right)\right) \cap \left(V^k \times V^k\right), \\
    \mathcal{E}^k &= \left(\mathcal{E}^{k - 1} \setminus \bigcup_{e \in \mathcal{E}_p^{k - 1}} \left(\mathcal{L}^{k - 1}_e \times \{e\}\right) \right) \cup \left(N^{k - 1}_p \times \{p\}\right),\\
    \mathcal{L}^k &= \left(\mathcal{L}^{k - 1} \setminus \bigcup_{e \in \mathcal{E}_p^{k - 1}} \left(\{e\} \times \mathcal{L}^{k - 1}_e\right) \right) \cup \left(\{p\} \times N^{k - 1}_p\right). 
\end{align*}

To build the minimum degree algorithm using quotient graphs, it is necessary to update the degrees of all variables $v \in N_p$ after eliminating the pivot $p$.
However, this step is computationally expensive because reconstructing the neighborhood of each $v$ requires performing set union operations, as described in Equation~(\ref{eq:invariant}).
These degree updates have been identified as the primary bottleneck of the minimum degree algorithm in the literature~\cite{liu1985modification,george1989evolution}.

\subsection{Multiple Minimum Degree Algorithm.}\label{sec:mmd}

The multiple minimum degree algorithm~\cite{liu1985modification} addresses the bottleneck in degree updates by selecting a maximal independent set of pivots, denoted by $M$, within an additive relaxation of the minimum degree criterion. 
All pivots in $M$ are eliminated sequentially with their associated connection updates applied before any degree updates are performed on affected variables.
The independence of these pivots ensures that pivots do not interfere with one another, while the maximality of the set increases the likelihood of overlapping pivots' neighborhoods.
This overlap is key to reducing the cost of degree updates: whereas the standard minimum degree algorithm requires a separate degree update for each adjacent pivot in $M$, the multiple minimum degree algorithm defers degree updates until all pivots in $M$ are eliminated, allowing each affected variable to perform only a single, consolidated degree update.

For example, consider selecting a maximal independent set $M = \{1, 3, 7, 9\}$ from $\mathcal{G}^0$ in Figure~\ref{fig:quotient}.
If the variables in $M$ are eliminated one by one using the minimum degree algorithm, then each of the variables 2, 4, 6, and 8 will incur two degree updates.
In contrast, with multiple elimination, where all variables in $M$ are eliminated in a single step, only one degree update is required for each of the variables 2, 4, 6, and 8 after all connection updates are applied.

\begin{figure}[tbp]
    \centering
    \includegraphics[width=\linewidth]{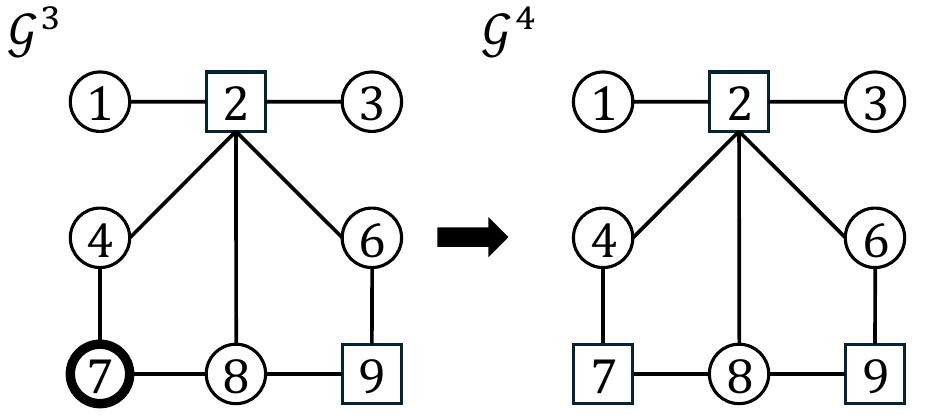}
    \vspace{-1em}
    \caption{An example illustrating how to compute approximate degrees after eliminating variable 7. To get the approximate degree of variable 8 in $\mathcal{G}^4$, we compute $|\mathcal{L}_7 \setminus \{8\}| + |\mathcal{L}_2 \setminus \mathcal{L}_7| + |\mathcal{L}_9 \setminus \mathcal{L}_7| = 5$.}
    \label{fig:approx_deg}
\end{figure}

\subsection{Approximate Minimum Degree Algorithm.}\label{sec:amd}
The approximate minimum degree (AMD) algorithm~\cite{amestoy1996approximate} adopts an alternative strategy to reduce the cost of degree updates.
Instead of computing exact degrees, the AMD algorithm estimates an upper bound, $d^k_v$, for the exact degree of a vertex $v$ in $G^k$:
\begin{align*}
    d^k_v = \min\left\{\hspace{-0.3em}\begin{array}{l}
          n - k - 1, \\
          d^{k - 1}_v + |\mathcal{L}_p \setminus \{v\}| - 1, \\
          |\mathcal{A}_v \setminus \{v\}| + |\mathcal{L}_p \setminus\{v\}| + \sum_{e \in \mathcal{E}_v} |\mathcal{L}_e \setminus \mathcal{L}_p|
    \end{array}\hspace{-0.3em}\right\}
\end{align*}
with $d^0_v$ initialized as the exact degree of $v$ in $G^0$. 
Note that $d_v$ remains unchanged if $v \notin \mathcal{L}_p$.
This estimate is known as the \textit{approximate degree}.
The first term reflects the maximum possible degree given the size of the remaining submatrix, the second accounts for the worst case fill-in introduced by $p$, and the third refines the estimate using union bound and avoids double counting using local information from the neighborhood of $p$.
An example is shown in Figure~\ref{fig:approx_deg}.

To efficiently compute $|\mathcal{L}_e \setminus \mathcal{L}_p|$, Algorithm~\ref{alg:approx} uses the identity $|\mathcal{L}_e \setminus \mathcal{L}_p| = |\mathcal{L}_e|
 - |\mathcal{L}_e \cap \mathcal{L}_p|$. 
 During the scan, if an element $e$ is encountered for the first time, $w(e)$ is set to $|\mathcal{L}_e|$, and then decremented once for every variable $v \in \mathcal{L}_e \cap \mathcal{L}_p$.
 Otherwise, $w(e)$ remains negative.
 The negativity assumption on $w(e)$ required by the algorithm is ensured using a timestamp method, as detailed in~\cite{amestoy1996approximate}.
 Combining both cases, the final value of $|\mathcal{L}_e \setminus \mathcal{L}_p|$ for each $e \in \overline{V}$ is given by~\cite{amestoy1996approximate}:
\begin{align*}
    |\mathcal{L}_e \setminus \mathcal{L}_p| = \left\{\begin{array}{ll}
        w(e) & \text{if } w(e) \ge 0,\\
        |\mathcal{L}_e| & \text{otherwise} 
     \end{array}\right\}.    
\end{align*}

\begin{algorithm}[htbp]
\caption{Computation of $|\mathcal{L}_e \setminus \mathcal{L}_p|$~\cite{amestoy1996approximate}}\label{alg:approx}
\begin{algorithmic}[1]
\State Assume $w(e) < 0$ for all $e \in \overline{V}$.
\For{each variable $v \in \mathcal{L}_p$} \label{line:intra}
\For{each element $e \in \mathcal{E}_v$}
    \If{$w(e) < 0$} \label{line:atomic_start}
    \State $w(e) \gets |\mathcal{L}_e|$ 
    \EndIf 
    \State $w(e) \gets w(e) - 1$ \label{line:atomic_end}
\EndFor
\EndFor
\end{algorithmic} 
\end{algorithm}

The AMD algorithm therefore selects, at each step, the variable with the smallest approximate degree in $\mathcal{G}^{k - 1}$ as the $k$th pivot.
For a detailed complexity analysis of both the AMD and the minimum degree algorithms, we refer the readers to~\cite{heggernes2002computational}.

The AMD algorithm also incorporates several additional techniques, including mass elimination~\cite{george1978application}, element absorption~\cite{duff1983multifrontal}, indistinguishable nodes~\cite{george1981computer}, and external degrees~\cite{liu1985modification}.
These techniques are included in our parallel implementation of the AMD algorithm, though we omit detailed discussion here, as they are orthogonal to parallelization and may be explored in the cited references for interested readers.

\subsection{Related Work.}

\subsubsection{Minimum Degree and its Variants.}

The minimum degree heuristic, a symmetric analog to Markowitz's algorithm~\cite{markowitz1957elimination}, was first proposed by Tinney and Walker~\cite{tinney1967direct}.
A graph-theoretical interpretation was later developed by Rose~\cite{rose1972graph}, who coined the term minimum degree algorithm.
Over the years, numerous enhancements have been proposed and have become standard components in modern implementations.
These include mass elimination~\cite{george1978application}, indistinguishable nodes~\cite{george1981computer}, quotient graphs~\cite{george1980fast} (also known as generalized element models or superelements~\cite{eisenstat1976applications, speelpenning1978generalized,  duff1983multifrontal}), incomplete degree updates~\cite{liu1985modification, eisenstat1982yale}, element absorption~\cite{duff1983multifrontal}, multiple elimination~\cite{liu1985modification}, and external degrees~\cite{liu1985modification}.
These techniques and their impacts were comprehensively summarized by George and Liu~\cite{george1989evolution}.

\subsubsection{Approximate Minimum Degree and its Variants.}
Since the degree updates have been identified as the primary bottleneck of the minimum degree algorithm in the literature~\cite{liu1985modification,george1989evolution}, several researchers~\cite{davis1997unsymmetric, davis1991unsymmetric, gilbert1992sparse, amestoy1996approximate} proposed computing upper bounds on the degrees rather than computing exact values.
Among these approaches, the AMD algorithm~\cite{amestoy1996approximate} has been shown to be the most effective.
Its sequential implementation in SuiteSparse~\cite{suitesparsegithub} is now widely used in modern sparse solvers.
The success of the AMD algorithm also spurred further interest in nonsymmetric matrices, leading to the development of the column approximate minimum degree algorithm~\cite{davis2004column}, which aims to reorder columns to minimize worst-case fill-in across all potential numerical row pivoting.
\subsubsection{Theoretical Development.}
The fill-in minimization problem was shown to be NP-complete by Yannakakis~\cite{yannakakis1981computing}.
Heggernes et al.~\cite{heggernes2002computational} analyzed the asymptotic complexity of the minimum degree and AMD algorithms.
Additionally, several theoretical developments have been made to understand the minimum degree algorithm and its variants more, including improvements to the minimum degree algorithm, randomized approaches, approximation approaches, and parameterized complexity~\cite{cummings2021fast, fahrbach2018graph,natanzon1998polynomial,bliznets2020lower,cao2020minimum,fomin2013subexponential}.

\subsubsection{Tie-breakings.}\label{sec:tie}
The need of tie-breaking strategies in the minimum degree algorithm and its variants has long been recognized by researchers~\cite{george1989evolution,duff2017direct,duff1976george,amestoy1996approximate,cavers1987tiebreaking}.
The ordering quality is remarkably sensitive to the initial ordering of the input: changes of the initial ordering can increase or decrease the number of fill-ins by up to a factor of three empirically~\cite{george1989evolution}.
Despite this sensitivity, an effective tie-breaking strategy remains elusive.
The state-of-the-art AMD implementation in SuiteSparse~\cite{suitesparsegithub} still suffers from tie-breaking limitations.
To decouple this issue when evaluating reordering algorithms, prior works~\cite{george1989evolution,amestoy1996approximate} resorted to randomly permuting the input matrix before applying reorderings.

\subsubsection{Parallelization.}
In contrast to nested dissection~\cite{george1973nested, gilbert1986analysis} where extensive efforts have been made to develop parallel implementations~\cite{chevalier2008pt,karypis1998parallel,lasalle2015efficient}, the AMD algorithm has seen little progress in this regard.
To our knowledge, only two parallelization attempts related to the AMD algorithm exist.
The first is in PT-Scotch~\cite{chevalier2008pt}, where parallel nested dissection is followed by a variant of the AMD algorithm~\cite{pellegrini2000hybridizing} independently within each partition; however, the AMD step itself remains sequential and this paradigm relies on the existence of good separators.
The second attempt is a parallelization of the column approximate minimum degree algorithm~\cite{chen1999toward}, in which the minimum degree criterion was removed to increase parallelism across an independent set of pivots.
Nevertheless, it resulted in poor scaling and did not report the quality of the resulting orderings.
Overall, to the best of our knowledge, there is currently no scalable parallel implementation of the AMD algorithm.

\subsubsection{Impact.}
While linear systems with multiple right-hand sides or identical sparsity patterns, but varying numerical values, may not gain significant benefits from our approach since the AMD ordering time can be amortized, there are applications where reordering cannot be reused. 
Specifically, this applies to systems with sparsity patterns that change over time. 
Examples include Incremental Potential Contact~\cite{li2020incremental} and adaptive remeshing~\cite{schmidt2023surface, sellan2020opening}. 
Moreover, the AMD algorithm has shown valuable not only for direct solvers but also for iterative solvers when generating preconditioners~\cite{tunnell2025empirical}.
\section{Parallel Approximate Minimum Degree Algorithm.}
In this section, we first explain why simply parallelizing the most computationally expensive component, approximate degree updates in Algorithm~\ref{alg:approx}, is insufficient for achieving a scalable implementation.
We then describe how we address this challenge by adapting the multiple elimination strategy~\cite{liu1985modification} to suit the approximate minimum degree (AMD) algorithm~\cite{amestoy1996approximate} using distance-2 independent sets.
By introducing this new parallel framework and minimizing contention among threads via specialized concurrent data structures tailored to the AMD algorithm, we present the first scalable shared memory AMD implementation.

\subsection{Intra-Elimination Parallelism.}
A simple approach to extract parallelism from Algorithm~\ref{alg:approx} is to parallelize by variables in line~\ref{line:intra} and use atomics for each access to $w(e)$ between line~\ref{line:atomic_start}  and line~\ref{line:atomic_end} within each elimination step.
However, this strategy does not scale.
As shown in
Table~\ref{tab:intra}, while the amount of parallelism, $|\mathcal{L}_p|$, appears reasonable, this method is not scalable since the amount of work, $\sum_{v \in \mathcal{L}_p}|\mathcal{E}_v|$, is not large enough compared to $|\mathcal{L}_p|$.
Moreover, memory contention is high because the number of unique elements accessed, $\left|\bigcup_{v\in \mathcal{L}_p}\mathcal{E}_v\right|$, is significantly smaller than the amount of parallelism, $|\mathcal{L}_p|$.
These together illustrate why simply parallelizing the approximate degree updates fails to deliver a scalable implementation.

\begin{table}[htbp]
  \vspace{-0.6em}
  \caption{Average sizes of the sets across all elimination steps, representing the amount of parallelism $|\mathcal{L}_p|$, the amount of work $\sum_{v \in \mathcal{L}_p}|\mathcal{E}_v|$, and the number of unique elements accessed $\left|\bigcup_{v \in \mathcal{L}_p}\mathcal{E}_v\right|$, by applying intra-elimination parallelism.}
  \label{tab:intra}
  \centering
  \begin{tabular}{l|ccc}
    \toprule
    Matrix & $|\mathcal{L}_p|$ & $\sum_{v \in \mathcal{L}_p}|\mathcal{E}_v|$ & $\left|\bigcup_{v \in \mathcal{L}_p}\mathcal{E}_v\right|$ \\
    \midrule
    \texttt{nd24k} & 329.7 & 587.5 & 14.0 \\
    \texttt{Flan\_1565} & 43.8 & 64.8 & 10.2 \\
    \texttt{nlpkkt240} & 80.5 & 542.8 & 56.3 \\
    \bottomrule
  \end{tabular}
  \vspace{-0.6em}
\end{table}

\subsection{Inter-Elimination Parallelism Exploiting Distance-2 Independent Sets.}\label{sec:inter}
Since extracting parallelism within a single elimination step has proven unscalable, we now turn to the possibility of parallelizing the AMD algorithm across elimination steps using multiple elimination~\cite{liu1985modification} discussed in Section~\ref{sec:mmd}.
While the multiple minimum degree algorithm~\cite{liu1985modification} improved sequential performance through multiple elimination, the AMD algorithm did not incorporate this technique~\cite{amestoy1996approximate}.
This is because the advantage of multiple elimination diminishes when degrees are approximated~\cite{amestoy1996approximate}.
However, multiple elimination provides an additional benefit: it enables parallelism across elimination steps by breaking dependencies between pivots.

\begin{figure}[t]
    \centering
    \includegraphics[width=\linewidth]{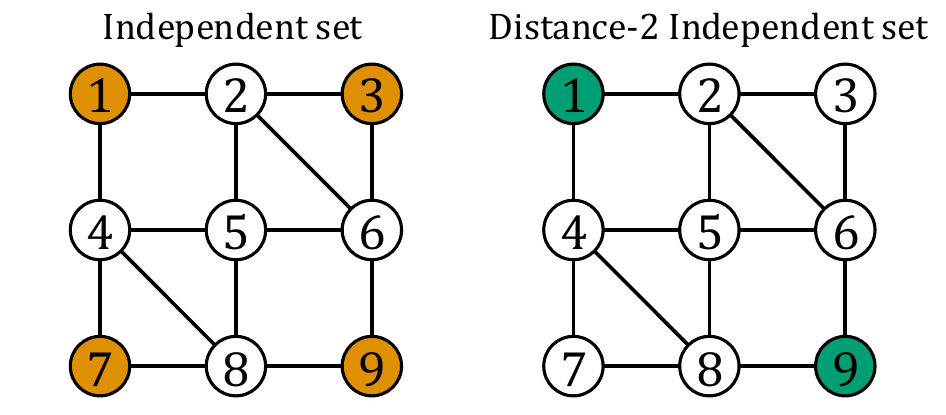}
    \vspace{-1.2em}
    \caption{When multiple elimination is applied to an independent set $\{1, 3, 7, 9\}$, memory contention arises because multiple pivots attempt to update the connections of variables 2, 4, 6, and 8. In contrast, using multiple elimination on a distance-2 independent set $\{1, 9\}$ avoids these issues entirely, where variables 2 and 4 are only adjacent to pivot 1 and variables 6 and 8 are only adjacent to pivot 9.}
    \label{fig:independent}
\end{figure}

To effectively leverage multiple elimination for parallelizing the AMD algorithm, we need to adapt the policy for selecting pivots. 
In the original design, performance gains from multiple elimination stemmed from the overlap of pivots' neighborhoods. 
However, such overlap is not well-suited to the AMD algorithm, as the degree of a variable is intended to be approximated with respect to the single adjacent pivot.
Moreover, overlapping neighborhoods introduce significant contention during connection updates, since a variable's updates can be affected by multiple pivots, even when those pivots do not directly interfere with each other.
To address these issues, we instead select a distance-2 independent set of pivots, ensuring no overlap in the pivots' neighborhoods as shown in Figure~\ref{fig:independent}.
Pivots in the distance-2 independent sets are then distributed among threads and processed independently in parallel.

We also relax the minimum approximate degree criterion for pivot selection by a multiplicative factor $\mathit{mult}$.
Specifically, we select a distance-2 independent set from the candidate variables whose approximate degrees are no greater than $\mathit{mult}$ times the current minimum approximate degree.
Table~\ref{tab:inter} shows the average sizes of maximal distance-2 independent sets across elimination steps for various values of $\mathit{mult}$, where all variables in the set are eliminated in each step.
The results show that relaxation is necessary to yield enough pivots to fully utilize all threads when pivots are processed in parallel.
However, overly aggressive relaxation can significantly degrade the ordering quality.
We will further analyze the trade-off between relaxation factor, ordering quality, and scalability in Section~\ref{sec:relaxation}.

\begin{table}[htbp]
  \caption{Average sizes of maximal distance-2 independent sets with various values of relaxation factor, $\mathit{mult}$.}
  \label{tab:inter}
  \centering
  \begin{tabular}{l|ccc}
    \toprule
    Matrix & $\mathit{mult}=1.0$ & $\mathit{mult}=1.1$ & $\mathit{mult}=1.2$ \\
    \midrule
    \texttt{nd24k} & 2.2 & 9.0 & 10.9 \\
    \texttt{Flan\_1565} & 42.0 & 448.5 & 678.1 \\
    \texttt{nlpkkt240} & 57.5 & 4084.5 & 6695.8 \\
    \bottomrule
  \end{tabular}
\end{table}

\subsection{Parallelizing the Core of the Approximate Minimum Degree Algorithm.}\label{sec:core}
Once multiple elimination via distance-2 independent sets is applied, most components of the AMD algorithm, including approximate degree updates and the other techniques mentioned in Section~\ref{sec:amd}, can be embarrassingly parallelized where each thread eliminates a pivot at a time until the distance-2 independent set is depleted.
However, two components still suffer from significant contention and dynamic memory management: connection updates and the maintenance of approximate degree lists.
In the following, we explain how we overcome these challenges.

\subsubsection{Concurrent Connection Updates.}
In the AMD implementation from SuiteSparse~\cite{suitesparsegithub}, the sets $\mathcal{A}, \mathcal{E}$ and $\mathcal{L}$ in the quotient graph are stored in a single compressed sparse column (CSC) structure with additional splitters to track the sizes of $\mathcal{A}_v$, $\mathcal{E}_v$ and $\mathcal{L}_e$ for each variable $v$ and element $e$.
As these adjacency sets can grow beyond their initial sizes during connection updates, extra space referred to as \textit{elbow room} is pre-allocated at the end of the CSC structure.
When an adjacency set expands, it acquires space from the elbow room and migrates its contents, leaving the old space unused.
Once the elbow room is exhausted, garbage collection is triggered to compact the CSC structure and reclaim unused memory, emptying elbow room at the end.

While it is possible to parallelize this structure using atomics to coordinate thread access to the elbow room, the approach suffers during garbage collection, as threads must synchronize and redo their ongoing updates after garbage collection is complete.
Instead, we observe empirically that the total extra memory required during elimination never exceeds 1.5 times the storage size of $E^0$.
This is primarily due to two facts: (i) for all variables $v$, the combined size $|\mathcal{A}_v| + |\mathcal{E}_v|$ never increases during elimination, and extra memory is only needed when creating $\mathcal{L}_p$ for a pivot $p$, whose size does not grow thereafter; and (ii) variables can be merged during elimination using techniques mentioned in Section~\ref{sec:amd}, which reduces the number of pivots and, consequently, the amount of extra memory required.
Therefore, by augmenting the CSC structure by 1.5 times in advance, we can empirically eliminate the need for garbage collection.
We acknowledge that some inputs may not follow this empirical observation, so we allow users to adjust the augmentation factor when necessary.
To further reduce contention, each thread performs only a single atomic operation to claim extra space after collecting all connection updates associated with the assigned pivots from the distance-2 independent set.
Notably, the use of a distance-2 independent set ensures that each variable is associated with at most one pivot, thereby minimizing contention.

\subsubsection{Concurrent Lists for Maintaining Approximate Degrees.}
In the AMD implementation from SuiteSparse~\cite{suitesparsegithub}, pivot selection is facilitated by maintaining $n$ degree lists. 
Each list corresponds to a specific approximate degree and stores the associated variables using a doubly linked list.
When a pivot is selected, it is removed from the list, and for each variable in its neighborhood, it is moved from its current list to the new one based on its updated approximate degree.
No dynamic memory allocation is required, as the total number of entries across all lists is bounded by $n$.
However, this data structure is not amenable to parallelization due to high contention when multiple threads attempt to access the same doubly linked list.

Instead, we let each thread $\mathit{tid}$ maintain its own structures: $n$ doubly linked lists $\mathit{dlist_{tid}}(0, \cdots, n - 1)$ to store variables based on their approximate degrees, a $\mathit{loc}$ array to record the local degree list to which each variable belongs, and an $\mathit{lamd}$ value to store the minimum approximate degree of variables locally maintained by this thread.
In addition, a shared $\mathit{affinity}$ array is used to track which thread holds the most up-to-date information for each variable.
Both the $\mathit{affinity}$ and $\mathit{loc}$ arrays are initialized to $-1$, and $\mathit{lamd}$ is initialized to $n$.
The use of a distance-2 independent set of pivots ensures that each variable is associated with at most one pivot---and thus with a single thread---during any given elimination step.

\begin{algorithm}[htbp]
\caption{Concurrent Approximate Degree Lists}\label{alg:list}
\begin{algorithmic}[1]
    \Function{Remove}{$tid, v$}
    \State $\mathit{affinity}(v) \gets -1$
    \EndFunction
    \State
    \Function{Insert}{$tid, v, deg$}
    \If{$loc(v) \neq -1$}
    \State $dlist_{tid}(loc(v)).remove(v)$
    \EndIf
    \State $dlist_{tid}(deg).insert(v)$
    \State $loc(v) \gets deg$
    \State $\mathit{affinity}(v) \gets tid$
    \State $lamd \gets \min\{deg, lamd\}$
    \EndFunction
    \State
    \Function{Get}{$tid, deg$}
    \For{$v \in dlist_{tid}(deg)$}
    \If{$\mathit{affinity}(v) \neq tid$}
    \State $dlist_{tid}(deg).remove(v)$
    \State $loc(v) \gets -1$
    \EndIf
    \EndFor
    \State \textbf{return} entries in $dlist_{tid}(deg)$
    \EndFunction
    \State
    \Function{Lamd}{$tid$}
    \While{$lamd < n$ and $\Call{Get}{tid, lamd} = \emptyset$}
    \State $lamd \gets lamd + 1$
    \EndWhile
    \State \textbf{return} $lamd$
    \EndFunction
\end{algorithmic}
\end{algorithm}

As shown in Algorithm~\ref{alg:list}, when a thread attempts to remove a variable $v$, it sets $\mathit{affinity}(v)$ to $-1$, effectively invalidating all entries of $v$ across all threads without modifying the lists.
When a thread attempts to insert a variable $v$, it first checks if $v$ is currently stored in the local degree lists based on $\mathit{loc}(v)$.
If so, i.e. $\mathit{loc}(v) \neq -1$, the stale entry of $v$ in the old degree list is explicitly removed.
In all cases, the variable $v$ is then inserted into the new degree list in thread $\mathit{tid}$, with $\mathit{loc}(v)$ set to the new approximate degree and $\mathit{affinity}(v)$ set to $\mathit{tid}$, indicating that the thread $\mathit{tid}$ has the latest information for variable $v$.
The value of $\mathit{lamd}$ is also updated accordingly.
Stale entries of variable $v$ stored in other threads, identified by a mismatch between $\mathit{affinity}(v)$ and the thread ID, are lazily reclaimed during list traversal for pivot selection using the $\Call{Get}{}$ function.
Since the value of $\mathit{lamd}$ is not updated when removing variables, an $\Call{Lamd}{}$ function is used to skip the empty degree lists and get the up-to-date minimum approximate degree maintained by a thread when requested.
This design avoids contention for degree list maintenance, with the only necessary coordination occurring when determining the global minimum approximate degree.

\begin{algorithm}[htbp]
\caption{Parallel Distance-2 Independent Set, an analog of Luby's algorithm~\cite{luby1985simple}}\label{alg:luby}
\begin{algorithmic}[1]
    \Function{Dist\_2\_Indep\_Set}{$mult, lim$}
    \State $amd \gets \min_{tid}(\Call{Lamd}{tid})$
    \ParFor{each thread $\mathit{tid}$}
    \State $Candidates \gets []$
    \For{integer $d \in [amd, \lfloor mult \times amd\rfloor$]}
        \State $Candidates.append(\Call{Get}{tid, d})$
        \If{$Candidates.size() \geq lim$}
        \State $Candidates.resize(lim)$
        \State \textbf{break}
        \EndIf
    \EndFor
    \For{variable $v \in Candidates$}
    \State $l(v) \gets (rand(), v)$
    \State $\forall u \in \{v\} \cup N_v, l_{min}(u) \gets (\infty, \infty)$
    \EndFor
    \State \textbf{barrier}
    \For{variable $v \in Candidates$}
    \State $\forall u \in \{v\} \cup N_v$, atomically update $l_{min}(u)$ 
    \State \quad by the minimum of $l_{min}(u)$ and $l(v)$
    \EndFor
    \State \textbf{barrier}
    \For{variable $v \in Candidates$}
    \If{$l(v) = l_{min}(u), \forall u \in \{v\} \cup N_v$}
    \State $v$ is a valid candidate
    \EndIf
    \EndFor
    \EndParFor
    \State \textbf{return} all valid candidates
    \EndFunction
\end{algorithmic} 
\end{algorithm}

\subsection{Parallel Distance-2 Independent Set.}
To parallelize the selection of a distance-2 independent set for multiple elimination, we first compute the global minimum approximate degree, $\mathit{amd}$, from the concurrent approximate degree lists.
We then collect candidate variables whose approximate degrees fall within the relaxed threshold described in Section~\ref{sec:inter}. This collection is performed in parallel from the local degree lists maintained by each thread.
To limit overhead compared to the sequential AMD algorithm, we restrict each thread to collect at most $\mathit{lim}$ candidates.
The impact of this parameter will be analyzed in Section~\ref{sec:relaxation}.

To select a distance-2 independent set, we perform a single iteration of the distance-2 analog of Luby's algorithm~\cite{luby1985simple} as shown in Algorithm~\ref{alg:luby}.
For each candidate gathered from the approximate degree lists, a random number is generated and compared against the candidates within its distance-2 neighborhood, with ties broken by candidate indices.
A candidate is considered valid if and only if it has the smallest value among the candidates within its distance-2 neighborhood.
To enable parallelism, Algorithm~\ref{alg:luby} includes appropriate atomic operations and synchronization barriers.
We intentionally avoid running the full algorithm because a maximal distance-2 independent set is unnecessary; as long as the set obtained in the first iteration is sufficiently large, it provides enough parallelism for the core AMD algorithm.

\subsection{Overall Parallel AMD Algorithm.}
Algorithm~\ref{alg:paramd} summarizes our strategies used to parallelize the AMD algorithm.
These techniques collectively enable an efficient and scalable implementation, addressing challenges in parallelizing the AMD algorithm.

\begin{algorithm}[htbp]
\caption{Parallel AMD Algorithm}\label{alg:paramd}
\begin{algorithmic}[1]
    \While{$|V| > 0$}
        \State $D \gets\Call{Dist\_2\_Indep\_Set}{mult, lim}$
        \ParFor{pivot $p$ in $D$}
            \State Eliminate $p$ using the core AMD algorithm
            \State \quad with concurrent connection updates and 
            \State \quad concurrent approximate degree lists
            \State \quad (see Section~\ref{sec:core})
        \EndParFor
    \EndWhile
\end{algorithmic} 
\end{algorithm}

\subsubsection{Memory Consumption.}
Let $n$ be the number of rows, $m$ the number of nonzeros in the input matrix, and $t$ the number of threads.
The AMD implementation from SuiteSparse~\cite{suitesparsegithub} requires $O(n + m)$ memory.
On top of that, our implementation allocates extra $1.5m$ memory for CSC augmentation, $O(nt)$ memory for the concurrent approximate degree lists, and $O(n)$ memory for distance-2 independent set selection.
All big-$O$ terms involve small constants, and in practice $nt$ is usually smaller than, or on the same order as, $m$; therefore, we expect our method to incur only a modest constant factor increase in memory.

\section{Experiments.}
\subsection{Evaluation Platforms.}
We evaluated our parallel AMD algorithm\footnote{\href{https://github.com/PASSIONLab/ParAMD}{\texttt{https://github.com/PASSIONLab/ParAMD}}} on an AMD EPYC 7763 CPU with 64 cores. 
The implementation was written in C++ using OpenMP and compiled with GCC 13.2.1 using the flags \texttt{-O3 -fopenmp}.
We set \texttt{OMP\_PROC\_BIND=close} so threads are in the same NUMA domain when possible.
As a baseline, we used the sequential implementation from SuiteSparse (v7.12.1)~\cite{suitesparsegithub}. 
To the best of our knowledge, no existing parallel implementation of the AMD algorithm is available for comparison.

\subsection{Matrix Suite.}
The sparse matrices used in our experiments were selected from the SuiteSparse Matrix Collection~\cite{davis2011university} and the M3E Matrix Collection~\cite{m3e}, covering a diverse range of large-scale linear systems arising from real-world applications. 
A summary of these matrices is provided in Table~\ref{tab:matrix}.

Note that the AMD algorithm remains a viable option for nonsymmetric solvers such as UMFPACK~\cite{davis2004algorithm} when the nonsymmetric matrix exhibits a nearly symmetric nonzero pattern. 
For this reason, we also included several nonsymmetric matrices in our dataset. 
In such cases, the AMD algorithm is applied to the matrix obtained by summing the absolute values of the original matrix and its transpose.
This pre-processing step is performed by SuiteSparse AMD regardless of whether the input matrix is symmetric.
For fair comparison, we parallelized this step using simple atomic operations and included its runtime in the reported results. 
However, users may skip this pre-processing step in our implementation when the input matrix is known to be symmetric.

\begin{table*}[htbp]
  \caption{A summary of the matrices used in our experiments, sorted by \#nonzeros.}
  \label{tab:matrix}
  \centering
  \begin{tabular}{l|cccc|l}
    \toprule
    Matrix & \#rows & \#nonzeros & Symmetric & Positive-definite & Description \\
    \midrule
    \texttt{nd24k} & 72.0K & 28.7M & $\checkmark$ & $\checkmark$ & 3D mesh problem\\
    \texttt{ldoor} & 952K & 42.5M & $\checkmark$ & $\checkmark$ & Structural problem \\
    \texttt{Serena} & 1.39M & 64.5M & $\checkmark$ & $\checkmark$ & Structural problem \\
    \texttt{dielFilterV3real} & 1.10M & 89.3M & $\checkmark$ & $\times$ & Electromagnetic problem \\
    \texttt{ML\_Geer} & 1.50M & 111M & $\times$ & $\times$ & Poroelastic problem \\
    \texttt{Flan\_1565} & 1.56M & 114M & $\checkmark$ & $\checkmark$ & Structural problem \\
    \texttt{Cube\_Coup\_dt0} & 2.16M & 124M & $\checkmark$ & $\times$ & Structural problem \\
    \texttt{Bump\_2911} & 2.91M & 128M & $\checkmark$ & $\checkmark$ & Geomechanical problem \\
    \texttt{Cube5317k} & 5.32M & 223M & $\checkmark$ & $\checkmark$ & Elasticity problem \\
    \texttt{HV15R} & 2.02M & 283M & $\times$ & $\times$ & Fluid dynamics problem \\
    \texttt{Queen\_4147} & 4.15M & 317M & $\checkmark$ & $\checkmark$ & Structural problem \\
    \texttt{stokes} & 11.4M & 349M & $\times$ & $\times$ & Semiconductor process problem \\
    \texttt{guenda11m} & 11.5M & 512M & $\checkmark$ & $\checkmark$ & Geomechanical problem \\
    \texttt{agg14m} & 14.1M & 633M & $\checkmark$ & $\checkmark$ & Mesoscale problem \\
    \texttt{rtanis44m} & 44.8M & 748M & $\checkmark$ & $\checkmark$ & Diffusion problem \\
    \texttt{nlpkkt240} & 28.0M & 761M & $\checkmark$ & $\times$ & Optimization problem \\
    \bottomrule
  \end{tabular}
\end{table*}

\subsection{Ordering Comparison.}
We now present the performance of our parallel AMD implementation compared to the SuiteSparse baseline, as summarized in Table~\ref{tab:end}.
We used the default relaxation factor $\mathit{mult} = 1.1$ and set the limitation factor $\mathit{lim}$ to $8192$ divided by the number of threads, following the heuristic that this configuration targets a fill-in ratio of $1.1\times$ over the SuiteSparse baseline while ensuring a sufficiently large pool of candidates is considered for the distance-2 independent set in each elimination step.

Table~\ref{tab:end} shows that our parallel implementation achieves up to a $7.29\times$ speedup over SuiteSparse using 64 threads.
Also, our fill-in ratio is consistently around $1.1\times$ over SuiteSparse, reflecting the effect of the relaxation factor $\mathit{mult}$ in avoiding the selection of overly suboptimal pivots.

\begin{table*}[htbp]
  \caption{Ordering comparison between SuiteSparse AMD and our 64-thread parallel AMD implementation. Matrices were randomly permuted five times with all methods evaluated on the same set of permutations to decouple tie-breaking issue explained in Section~\ref{sec:tie}. Ordering time is reported as mean $\pm$ standard deviation due to its variability across runs, while the fill-in results are more consistent where the standard deviation is at most 6\% of the mean so they are reported as mean only over the five runs. Speedup is computed per run and we report the mean speedup across five runs.}
  \label{tab:end}
  \centering
  \begin{tabular}{l|cc|c|cc|c}
    \toprule
    \multirow{3}{*}{Matrix} & \multicolumn{2}{c|}{Ordering Time (sec)} & \multicolumn{1}{c|}{Our 64-thread} & \multicolumn{2}{c|}{\#Fill-ins} & \multicolumn{1}{c}{Our}\\\cline{2-3}\cline{5-6}
    & \multirow{2}{*}{SuiteSparse} & \multirow{2}{*}{Ours} & \multicolumn{1}{c|}{Speedup over} & \multirow{2}{*}{SuiteSparse} & \multirow{2}{*}{Ours} & \multicolumn{1}{c}{Fill-in} \\
    & & & SuiteSparse & & & \multicolumn{1}{c}{Ratio} \\
    \midrule
    \texttt{nd24k} & 0.82 $\pm$ 0.00 & 0.26 $\pm$ 0.00 & (3.18 $\pm$ 0.04)$\times$ & 5.03e+08 & 5.18e+08 & 1.02$\times$ \\
    \texttt{ldoor} & 1.42 $\pm$ 0.05 & 0.32 $\pm$ 0.03 & (4.39 $\pm$ 0.24)$\times$ & 1.52e+08 & 1.57e+08 & 1.04$\times$ \\
    \texttt{Serena} & 2.81 $\pm$ 0.17 & 0.56 $\pm$ 0.05 & (5.06 $\pm$ 0.20)$\times$ & 7.48e+09 & 8.79e+09 & 1.19$\times$ \\
    \texttt{dielFilterV3real} & 2.96 $\pm$ 0.33 & 0.54 $\pm$ 0.12 & (5.59 $\pm$ 0.54)$\times$ & 9.60e+08 & 1.11e+09 & 1.16$\times$ \\
    \texttt{ML\_Geer} & 4.26 $\pm$ 0.48 & 0.77 $\pm$ 0.20 & (5.71 $\pm$ 0.77)$\times$ & 1.42e+09 & 1.46e+09 & 1.03$\times$ \\
    \texttt{Flan\_1565} & 4.85 $\pm$ 0.56 & 0.95 $\pm$ 0.27 & (5.31 $\pm$ 0.80)$\times$ & 3.71e+09 & 3.94e+09 & 1.06$\times$ \\
    \texttt{Cube\_Coup\_dt0} & 5.21 $\pm$ 0.65 & 1.15 $\pm$ 0.30 & (4.65 $\pm$ 0.58)$\times$ & 2.24e+10 & 2.52e+10 & 1.15$\times$ \\
    \texttt{Bump\_2911} & 6.78 $\pm$ 0.74 & 1.52 $\pm$ 0.50 & (4.76 $\pm$ 0.91)$\times$ & 4.27e+10 & 4.83e+10 & 1.15$\times$ \\
    \texttt{Cube5317k} & 13.94 $\pm$ 0.92 & 2.90 $\pm$ 0.67 & (4.99 $\pm$ 0.89)$\times$ & 6.29e+09 & 6.54e+09 & 1.04$\times$ \\
    \texttt{HV15R} & 14.43 $\pm$ 1.31 & 2.34 $\pm$ 0.33 & (6.23 $\pm$ 0.66)$\times$ & 5.57e+10 & 5.63e+10 & 1.01$\times$ \\
    \texttt{Queen\_4147} & 15.45 $\pm$ 1.81 & 2.87 $\pm$ 0.79 & (5.61 $\pm$ 0.82)$\times$ & 7.70e+10 & 8.43e+10 & 1.10$\times$ \\
    \texttt{stokes} & 85.35 $\pm$ 6.51 & 14.31 $\pm$ 4.81 & (6.47 $\pm$ 1.52)$\times$ & 7.79e+10 & 8.17e+10 & 1.06$\times$ \\
    \texttt{guenda11m} & 36.59 $\pm$ 2.55 & 8.53 $\pm$ 2.47 & (4.64 $\pm$ 1.27)$\times$ & 3.10e+11 & 3.68e+11 & 1.17$\times$ \\
    \texttt{agg14m} & 56.10 $\pm$ 9.38 & 10.63 $\pm$ 2.99 & (5.62 $\pm$ 1.38)$\times$ & 3.75e+11 & 4.04e+11 & 1.09$\times$ \\
    \texttt{rtanis44m} & 195.56 $\pm$ 26.25 & 42.76 $\pm$ 12.06 & (4.80 $\pm$ 0.92)$\times$ & 1.12e+12 & 1.25e+12 & 1.13$\times$ \\
    \texttt{nlpkkt240} & 393.36 $\pm$ 57.62 & 56.31 $\pm$ 15.18 & (7.29 $\pm$ 1.45)$\times$ & 1.60e+12 & 1.82e+12 & 1.14$\times$ \\
    \bottomrule
  \end{tabular}
\end{table*}

\subsection{Runtime Breakdown and Parallelism Analysis.}
\begin{figure*}[htbp]
    \centering
    \includegraphics[width=\linewidth]{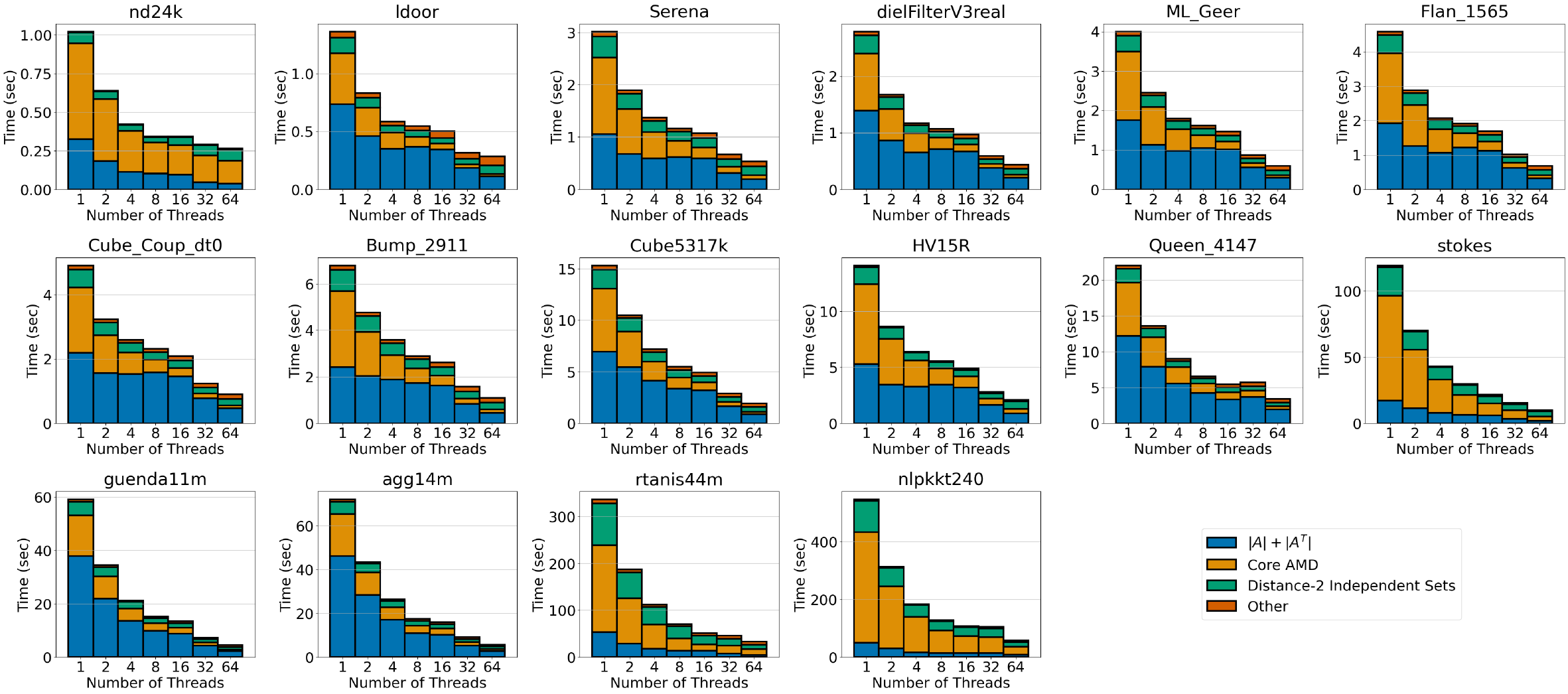}
    \vspace{-1em}
    \caption{Runtime breakdown of our parallel AMD algorithm as the number of threads scales from 1 to 64.}
    \label{fig:breakdown}
\end{figure*}

To gain deeper insights into our parallel AMD implementation, we present a runtime breakdown in Figure~\ref{fig:breakdown} using the default parameters.

First, when running with a single thread, our implementation is slower than the baseline.
This is primarily due to the inclusion of Algorithm~\ref{alg:luby} for selecting distance-2 independent sets, a step not required in the original AMD algorithm.
However, it is this modification that enables effective parallelization that makes our implementation scalable.

Second, we observe that the pre-processing step, which computes $|A| + |A^T|$, becomes a bottleneck for some matrices and does not scale well with increasing thread count. 
This behavior is expected due to high contention and irregular memory access inherent to the operation. 
Therefore, we suggest users skip this pre-processing step if the input is known to be symmetric.

\begin{figure*}[htbp]
    \centering
    \includegraphics[width=\linewidth]{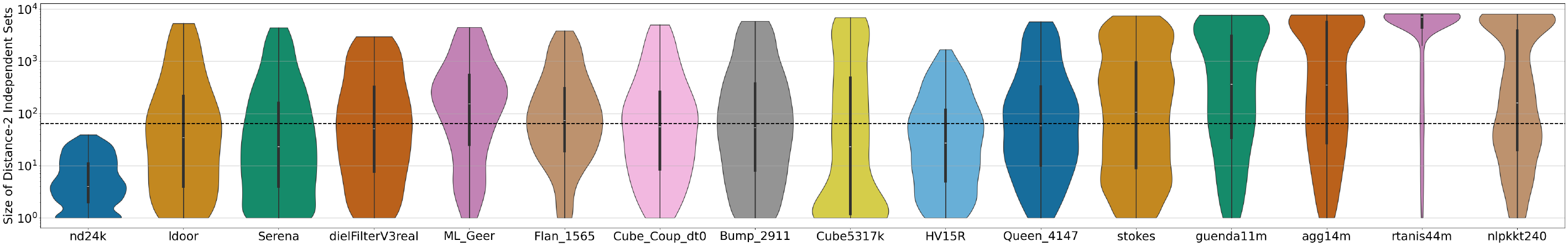}
    \vspace{-1em}
    \caption{Violin plots showing the distribution of the sizes of distance-2 independent sets across elimination steps. The plots are cropped to reflect the actual range of the data. The dotted line marks the threshold of 64, representing the minimum size needed to fully utilize all 64 threads.}
    \label{fig:violin}
\end{figure*}

Third, while the core AMD part is highly parallelizable in theory, it does not achieve perfect scaling in practice. 
This limited scalability is explained by the distribution of the sizes of the distance-2 independent sets, as shown in Figure~\ref{fig:violin}.
Notably, for the matrix that scales the worst, \texttt{nd24k}, the distance-2 independent sets selected are relatively small compared to other matrices, resulting in underutilization of threads. 
Moreover, across all matrices, a significant portion of the distance-2 independent sets still have sizes less than 64, further illustrating why the core AMD algorithm does not scale perfectly despite being embarrassingly parallel.

\begin{figure*}[htbp]
    \centering
    \includegraphics[width=0.9\linewidth]{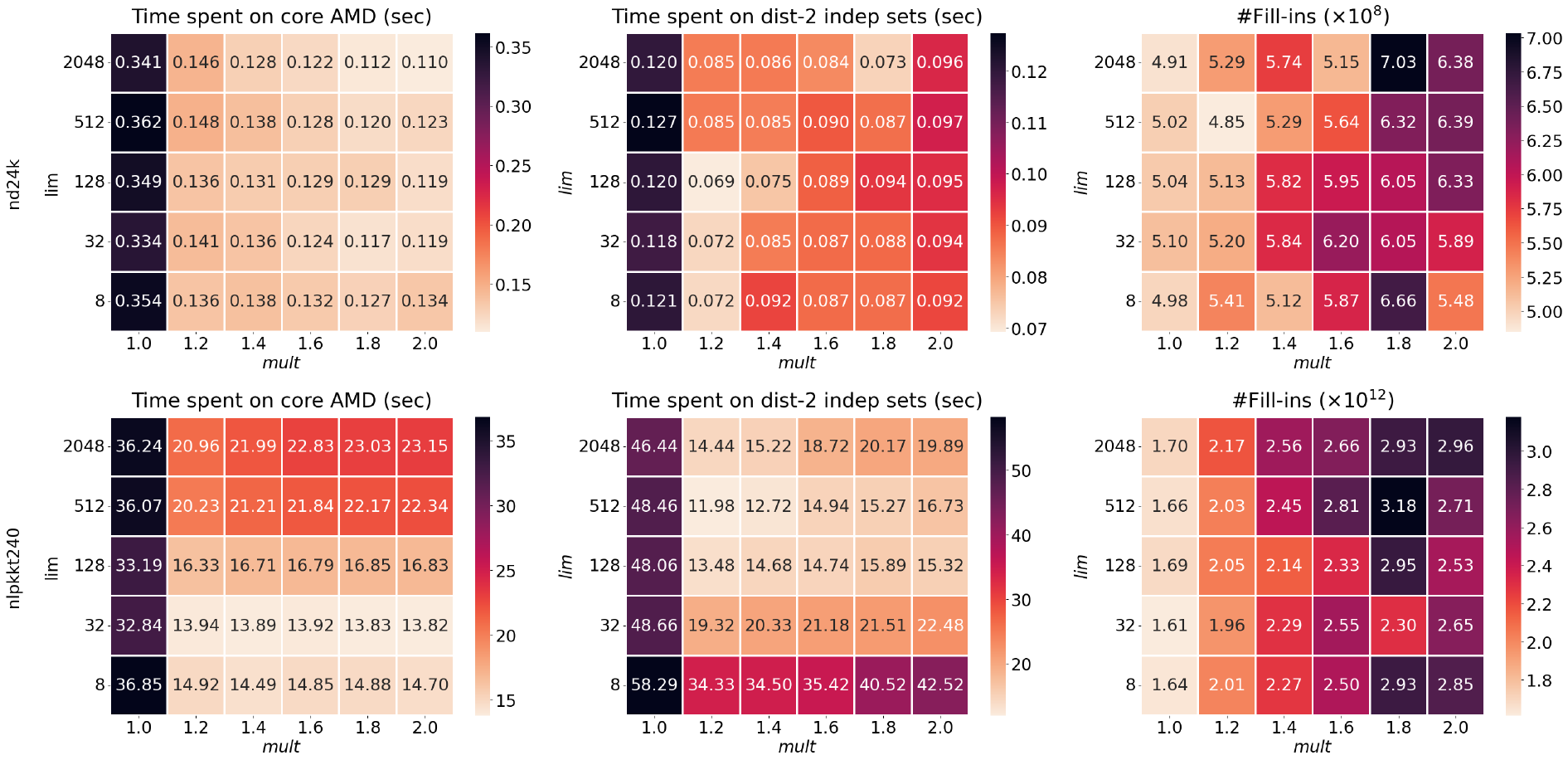}
    \vspace{-1em}
    \caption{Impact of the relaxation factor $mult$ and the limitation factor $lim$. Results are shown for matrices \texttt{nd24k} and \texttt{nlpkkt240}, representing the worst and best scalability cases, respectively. All experiments were conducted using 64 threads. Lighter colors indicate better performance.}
    \label{fig:heat}
\end{figure*}

\subsection{Relaxation and Limitation Factors.}\label{sec:relaxation}
We now evaluate the impact of the relaxation factor $\mathit{mult}$ and the limitation factor $\mathit{lim}$.
Figure~\ref{fig:heat} shows how they affect the core AMD runtime, the runtime of distance-2 independent set selection, and the ordering quality. 
Our analysis focuses on two representative matrices using 64 threads: \texttt{nd24k} and \texttt{nlpkkt240}, which demonstrates the worst and the best scalability, respectively.

The impact of the relaxation factor $\mathit{mult}$ is more pronounced and consistent. 
When $\mathit{mult}$ is too small, the algorithm lacks sufficient parallelism, leading to poor performance. 
On the other hand, excessively large values of $\mathit{mult}$ degrade the ordering quality due to the inclusion of suboptimal pivots.

In contrast, the impact of the limitation factor $\mathit{lim}$ is less straightforward. 
This is because adjusting $\mathit{lim}$ affects several aspects of the algorithm, including the number of elimination steps, opportunities to apply other techniques discussed in Section~\ref{sec:amd}, and the amount of computation available to hide memory access latency. 
As these factors impact different matrices to varying degrees, the limitation factor $\mathit{lim}$ produces complex behavior with no single dominant pattern.

Nevertheless, Figure~\ref{fig:heat} suggests that the optimal configuration occurs around $\mathit{mult} = 1.2$ and $\mathit{lim} = 128$. 
After further fine-grained tuning across our dataset, we found that setting $\mathit{mult} = 1.1$ provides a better balance between runtime and ordering quality. 
To ensure that the number of candidates for the distance-2 independent set remains constant regardless of the number of threads, we finalize $\mathit{lim}$ as 8192 divided by the number of threads. 
This results in $\mathit{lim} = 128$ when using 64 threads, aligning with the identified optimal value.
These together result in the default settings.

\begin{table*}[htbp]
  \caption{End-to-end comparison of ordering methods for solving sparse symmetric positive definite systems. Each matrix is randomly permuted five times, with all methods evaluated on the same set of permutations to avoid the tie-breaking issues discussed in Section 2.5.4. All orderings are computed on the CPU---SuiteSparse AMD uses single thread and the others use 64 threads---while the reordered systems are solved on the GPU using cuDSS in double precision.}
  \label{tab:overall}
  \centering
  \begin{tabular}{l|cc|cc|cc}
    \toprule
    \multirow{3}{*}{Matrix} & \multicolumn{2}{c|}{SuiteSparse AMD} & \multicolumn{2}{c|}{Ours} & \multicolumn{2}{c}{cuDSS ND}\\\cline{2-7}
    & Ordering & GPU Solver & Ordering & GPU Solver & Ordering & GPU Solver \\
    & Time (sec) & Time (sec) & Time (sec) & Time (sec) & Time (sec) & Time (sec) \\
    \midrule
    \texttt{nd24k} & 0.82 $\pm$ 0.00 & 1.97 $\pm$ 0.07 & 0.26 $\pm$ 0.00 & 1.97 $\pm$ 0.09 & 2.38 $\pm$ 0.13 & 1.27 $\pm$ 0.02 \\
    \texttt{ldoor} & 1.42 $\pm$ 0.05 & 3.03 $\pm$ 0.01 & 0.32 $\pm$ 0.03 & 3.07 $\pm$ 0.01 & 0.67 $\pm$ 0.01 & 2.48 $\pm$ 0.01 \\
    \texttt{Flan\_1565} & 4.85 $\pm$ 0.56 & 18.92 $\pm$ 0.22 & 0.95 $\pm$ 0.27 & 19.62 $\pm$ 0.23 & 2.41 $\pm$ 0.01 & 9.21 $\pm$ 0.02 \\
    \texttt{Cube5317k} & 13.94 $\pm$ 0.92 & 43.90 $\pm$ 0.37 & 2.90 $\pm$ 0.67 & 44.33 $\pm$ 0.41 & 7.43 $\pm$ 0.04 & 33.45 $\pm$ 0.21 \\
    \bottomrule
  \end{tabular}
  \vspace{-1em}
\end{table*}

\subsection{End-to-End Solver Performance.}
Lastly, we evaluated end-to-end solver performance by solving the reordered systems using orderings computed by SuiteSparse AMD, our parallel AMD implementation, and a multi-threaded nested dissection (ND) method. 
For the solver, we focused on symmetric positive definite systems---eliminating the need for numerical pivoting---and employed cuDSS~\cite{cudss} (v0.7.1) on an NVIDIA A100 80GB GPU in double precision.
The multi-threaded ND used in our evaluation is the one that comes with cuDSS, which is a customized variant of METIS~\cite{karypis1997metis}.
Note that all orderings are computed on the CPU---SuiteSparse AMD uses single thread and the others use 64 threads---while the reordered systems are solved on the GPU.
For this experiment, we restricted our evaluation to only four matrices in our dataset as the input matrices need to be symmetric positive definite and their Cholesky factors need to fit within the available GPU memory.

From Table \ref{tab:overall}, we observe that the end-to-end performance improves with our parallel AMD compared to SuiteSparse AMD.
While GPU running time increases slightly due to extra fill-ins, this overhead is outweighed by the reduced ordering time.
Although the extra fill-ins are generally not problematic, we acknowledge that they may cause out-of-memory issues in some cases.
Taking the \texttt{Serena} matrix as an example, its Cholesky factor barely fits on the GPU when using SuiteSparse AMD, but exceeds the limit when using our parallel AMD.
Therefore, it is omitted here.
On the other hand, our parallel AMD outperforms multi-threaded ND for the reordering phase, with the slowdown occurring in the GPU solver time.
Systems reordered by multi-threaded ND can be solved faster because they have fewer fill-ins, as summarized in Table \ref{tab:fill}.
Moreover, the solver benefits from the partitions produced by ND, which enable better parallelism during factorization.
We believe a hybrid approach combining our method with multi-threaded ND could yield a new parallel ordering that further enhances end-to-end solver performance.

\begin{table}[htbp]
  \vspace{-0.5em}
  \caption{Comparison of the \#Fill-ins by using the orderings from SuiteSparse AMD, our parallel AMD, and cuDSS ND.}
  \vspace{0.2em}
  \label{tab:fill}
  \centering
  \begin{tabular}{l|ccc}
    \toprule
    \multirow{3}{*}{Matrix} & \multicolumn{3}{c}{\#Fill-ins} \\\cline{2-4}
     & SuiteSparse & \multirow{2}{*}{Ours} & cuDSS \\
     & AMD & & ND \\
    \midrule
    \texttt{nd24k} & 5.03e+08 & 5.18e+08 & 3.21e+08 \\
    \texttt{ldoor} & 1.52e+08 & 1.57e+08 & 1.41e+08 \\
    \texttt{Flan\_1565} & 3.71e+09 & 3.94e+09 & 1.46e+09 \\
    \texttt{Cube5317k} & 6.29e+09 & 6.54e+09 & 4.02e+09 \\
    \bottomrule
  \end{tabular}
  \vspace{-1.2em}
\end{table}
\section{Conclusion and Future Work.}
We have presented the first scalable shared memory AMD implementation.
By incorporating distance-2 independent sets into multiple elimination and designing specialized concurrent
data structures tailored to the AMD algorithm, we have enabled effective parallelism and reduced contention.
Our experiments have demonstrated that this approach scales and consistently outperforms the AMD implementation from SuiteSparse, with only modest and controllable overhead in fill-in.

As immediate future work, we plan to further investigate the impact of the limitation factor and explore strategies for dynamically adapting both the relaxation and limitation factors during elimination, particularly when low workload is detected, to extract greater parallelism when needed.
In the longer term, we aim to address tie-breaking issues and develop an efficient, parallel-friendly approach for this critical component.
\section*{Acknowledgments.}
This research is supported by the Applied Mathematics program of the Advanced Scientific Computing Research (ASCR) within the Office of Science of the DOE under Award Number DE-AC02-05CH11231. 
We used resources of the National Energy Research Scientific Computing Center (NERSC), a Department of Energy Office of Science User Facility using NERSC award ASCR-ERCAP-33069.
\newpage
\bibliographystyle{siamplain}
\bibliography{references}
\end{document}